\documentclass[epj,draft]{svjour}

\usepackage{latexsym}
\usepackage{graphics}
\usepackage{amsmath}   
\usepackage{etoolbox}
\usepackage{cite}
\usepackage{hyperref}   % use for hypertext links, including those to external documents and URLs

\newcommand{\fant}[1]{\phantom{#1}}
\newcommand{\be}{\begin{equation}}
\newcommand{\ee}{\end{equation}}
\newcommand{\wdg}{\wedge}
\newcommand{\ot}{\otimes}

\begin{document}
\title{A derivation of Weyl-Lanczos equations}
\author{Ahmet BAYKAL% etc
\thanks{abaykal@ohu.edu.tr}%
\and Burak \"UNAL%
\thanks{burakunal90@gmail.com}
}
\institute{Department of Physics, 
Faculty of Arts and Sciences, 
Ni\u gde \"Omer Halisdemir University,  
Merkez Yerle\c{s}ke, Bor yolu \"uzeri, 
51240   Ni\u gde, TURKEY}
\date{10 January 2018}
\abstract{
The Lanczos potential for the Weyl tensor is derived from a quadratic curvature Lagrangian by making use of the exterior algebra of forms and the variational principles with constraints.
 % end of PACS codes
} 
\maketitle
\section{Introduction}

The Einstein's field equations determine the geometry of the spacetime by determining the Ricci tensor in terms of energy-momentum distribution of matter 
in a spacetime. On the other hand, as a consequence of the Einstein's equivalence principle, gravity can locally be transformed away. The corresponding mathematical statement is that the Christoffel symbols $\Gamma^{\mu}_{\alpha\beta}$ can be made to vanish at a point by transforming into a suitable coordinate system.  
But the corresponding Riemann tensor cannot be transformed away by any coordinate transformation, and the geometrical nature of gravity can  be observed  as relative accelerations of point particles following nearby geodesic curves, and the relative accelerations are determined by the geodesic deviation equation
involving the full Riemann tensor but not only its contractions. (See, for example, Ref. \cite{straumann}.) Conversely, it is possible to determine the Riemann tensor locally  from the measurements of geodesic deviation  \cite{szekeres}. Therefore, the observable  effect of a distant gravitating mass on the motion of a test particle is encoded in the  Weyl tensor on the geodesic curve followed by the particle in vacuum. In this regard, it is appropriate to note that the Weyl curvature tensor, occurring for the first time in four spacetime dimensions, represents the  free gravitational field that can  experimentally be determined by measurements of the tidal forces.   However, the fourth and the second rank irreducible parts of a Riemann tensor are not independent of one  another since they are related by the second Bianchi identity. The same identity can be used to derive the geodesic postulate for the test particles as well.  The current discussion is about a third rank tensor, namely the Lanczos spintensor from which the Weyl tensor can locally be derived in parallel to the case of a gauge field strength derived from a gauge potential.

While investigating the self-dual part of the Riemann tensor,  Lanczos  discovered \cite{cl_0,cl_1,cl_2,takeno}, 
 that Weyl tensor can be expressed in terms of a third order tensor potential. In addition,
although the algebraic classification  of the Weyl tensor by Petrov \cite{petrov} was introduced after the Lanczos tensor had been introduced, 
\cite{penrose,NP}, it took some time for the Lanczos potential theory to be discussed by using the 2-spinor formalism by which the classification 
of the Weyl tensor can  be concisely presented. Relatively recently, Zund and Maher \cite{Maher1968}, and Taub \cite{TAUB1975377}  developed the spinorial version of the Lanczos potential. The construction of the Lanczos potential  with respect  to an orthonormal as well as a null coframe can be of considerable 
computational use as will be illustrated below. 

Lanczos' himself argued \cite{cl_1} that a tensor potential for the Weyl tensor is unique to four spacetime dimensions as proved by Edgar and H\"oglund
\cite{Edgar2000}. On the other hand, Edgar proved that a tensor potential for the Weyl tensor for higher dimensions does not exist \cite{Edgar1994}.
More recently, Edgar and Senovilla  \cite{Edgar2004} showed that in dimensions at $n\geq 4$ there is a new type of tensor potential 
of the form $P^{ab}_{\fant{ab}cde}$ whereas  at four dimensions the Lanczos' potential $H^a_{\fant{a}bc}$ 
corresponds to the dual of the general potential $P^{ab}_{\fant{ab}cde}$. For a more accurate historical account of the discovery of, what is now commonly called, the Lanczos spintensor (or Lanczos tensor potential), as well as for a current list of important works in the literature, the reader is referred to the review article \cite{odonnell} by O'Donnell. The discussion below is confined to four spacetime dimensions and the construction is independent of the metric signature.

The paper is organized as follows. The notation used and the related mathematical definitions  are 
introduced in Sec.\ref{defs-sec}. The irreducible parts of the Riemann curvature tensor, their duality relations and the inner products of these parts are also discussed in Sec.\ref{defs-sec}. All relations playing important role in the discussion are derived from scratch to make the presentation as much self-contained as possible. The derivations of duality relations for the irreducible parts of the Riemann tensor are  presented in the appendix. In Sec.
\ref{L_Lagrangian+Constraints}, the Lanczos' quadratic curvature Lagrangian and two additional constraints are introduced. The Lagrange multiplier terms for the constraints are then expressed in terms of tensor-valued differential forms and a compact expression for the Weyl-Lanczos equations is derived  by a variational procedure by eliminating one of the constraints. The expression obtained for the Weyl tensor  is then compared with the previous expressions  in the literature in Sec.\ref{comparison}. The special gauge invariance related to the  extended  Lanczos Lagrangian and the use of the gauge conditions on the Lanczos potential are discussed in Sec.\ref{gauge_invariance}. As an application of the developed calculation technique, a Lanczos potential is explicitly calculated for a type N plane gravitational wave metric using a complex null coframe in Sec.\ref{Lanczos-pot-for-pp-waves}. The refutation of the claim that the Riemann tensor can be derived from a tensor potential in the form presented originally by Massa and Pagani \cite{massa} is  discussed in Sec.\ref{massa-sec}. The presentation ends with a brief concluding  section pointing out some further applications.

\section{Definitions and notation}\label{defs-sec}

In this section, preliminary definitions of the geometrical quantities required in the construction of the Lanczos potential for the Weyl tensor
are presented. A more thorough and rigorous treatment of the tensor-valued differential forms can be found, for example, in 
the book  by Straumann \cite{straumann}. The notation used here follows closely the notation used by Straumann.

In terms of an orthonormal coframe $\{\theta^a\}$ with the range $a=0, 1, 2, 3$ related to a coordinate coframe by $\theta^a=e^{a}_{\mu}dx^\mu$, the metric tensor can be written in the form
\be
g
=
\eta_{ab}\theta^a\ot \theta^b
=
g_{\mu\nu}dx^\mu\ot dx^\nu
\ee
with constant components $\eta_{ab}=diag(-+++)$  and  the coordinate components can be expressed in terms of  $e^{a}_{\mu}$ as
\be
g_{\mu\nu}=e^{a}_{\mu}e^{b}_{\nu}\eta_{ab}.
\ee
where $e^{a}_{\mu}e^{\mu}_{b}=\delta^{a}_{b}$ and $e^{a}_{\mu}e^{\nu}_{a}=\delta^{\nu}_{\mu}$.
The Levi-Civita connection 1-forms $\omega^{a}_{\fant{a}b}=\omega^{a}_{cb}\theta^c$ satisfy the Cartan's first structure equations with zero torsion as
\be\label{se1}
d\theta^a+\omega^{a}_{\fant{a}b}\wdg \theta^b=0,
\ee
whereas the vanishing nonmetricity constraint for a Levi-Civita connection  reads $\omega_{ab}=\omega_{[ab]}$.
In terms of the connection 1-forms, the covariant derivative of basis coframe $\theta^a$ with respect to a vector field $V=V^ae_a$ can be expressed 
$\nabla_V\theta^a=-i_V(\omega^{a}_{\fant{a}b})\theta^b$. $D$ stands for the covariant exterior derivative of the Levi-Civita connection acting on tensor-valued forms.
Interior product with respect to a vector field $V$ is denoted by $i_V=V^ai_a$ and for the interior product with respect to a basis   frame field 
$e_a=e^{\mu}_a\partial_\mu$, the symbol $i_a\equiv i_{e_a}$ is used.

The components of the Riemann tensor $R^{ab}_{\fant{ab}cd}$ can be obtained from the expression for curvature 2-form
\be
\Omega^{ab}
=
\frac{1}{2}R^{ab}_{\fant{ab}cd}\theta^{c}\wdg \theta^{d}
\ee
and the curvature 2-form can be expressed in terms of Levi-Civita connection 1-forms $\omega^{a}_{\fant{a}b}$ as 
\be\label{se2}
\Omega^{ab}
=
d\omega^{ab}
+
\omega^{a}_{\fant{a}c}\wdg\omega^{cb}.
\ee
The Ricci 1-form and the scalar curvature can be defined in terms of the contraction of curvature 2-form
as $R^b=i_a\Omega^{ab}$ and $R=i_bi_a\Omega^{ab}$, respectively. 

Acting on basis coframe 1-forms, a right dual of the unity defines the oriented, invariant volume 4-form 
\be
*1
=\frac{1}{4!}\epsilon_{abcd}\theta^{abcd}=\sqrt{|g|}\varepsilon_{\mu\nu\alpha\beta}dx^\mu\wdg dx^\nu\wdg dx^\alpha\wdg dx^\beta
\ee
The indexes of the permutation symbol $\epsilon_{abcd}$ with $\epsilon_{0123}=+1$ are raised and lowered by the flat metric $\eta^{ab}$ and $\eta_{ab}$, respectively.
In contrast, the indexes of $\varepsilon_{\mu\nu\alpha\beta}$ are raised and lowered by the metric  components $g^{\mu\nu}$ and $g_{\mu\nu}$.
The exterior product of the coframe basis $p$-form is abbreviated as 
\be
\theta^{a_1a_2\cdots a_p}
\equiv
\theta^{a_1}\wdg \theta^{a_2}\wdg \cdots\wdg \theta^{a_p}
=
p!\theta^{[a_1}\ot \theta^{a_2}\ot \cdots \ot \theta^{a_p]}
\ee
where the square (round) brackets denote antisymmetrization (symmetrization) of the indexes enclosed.

The following definitions of left (Hodge) and right duals for curvature 2-forms are to be used extensively in what follows: 
\begin{align}
*\Omega^{ab}
&=
\frac{1}{2}R^{ab}_{\fant{ab}cd}*\theta^{cd}
=
\frac{1}{4}R^{ab}_{\fant{ab}mn}\epsilon^{mn}_{\fant{mn}cd}\theta^{cd},
\label{duality-defs1}\\
\Omega^{ab}*
&=
\frac{1}{2}\epsilon^{ab}_{\fant{ab}cd}\Omega^{cd}
=
\frac{1}{4}\epsilon^{ab}_{\fant{ab}mn}R^{mn}_{\fant{mn}cd}\theta^{cd},
\label{duality-defs2}\\
\tilde{\Omega}^{ab}
\equiv
*\Omega^{ab}*
&=
\frac{1}{2}\epsilon^{ab}_{\fant{ab}cd}*\Omega^{cd}
=
\frac{1}{8}\epsilon^{ab}_{\fant{ab}mn}R^{mn}_{\fant{mn}rs}\epsilon^{rs}_{\fant{ab}cd}\theta^{cd}.\label{duality-defs3}
\end{align}
The components of the left-, right- and double-dual Riemann tensors are given by the expression on the right hand side of the above expressions, respectively. 
For example, with $\tilde{\Omega}^{ab}=\frac{1}{2}\tilde{R}^{ab}_{\fant{ab}cd}\theta^{cd}$, from (\ref{duality-defs3}) one finds
\be\label{dd-curv-comp}
\tilde{R}^{ab}_{\fant{ab}cd}
=
\frac{1}{4}\epsilon^{ab}_{\fant{ab}rs}\epsilon^{pq}_{\fant{mn}cd}R^{rs}_{\fant{ab}pq}
=
-\frac{1}{4}\delta^{abpq}_{cdrs}R^{rs}_{\fant{ab}pq},
\ee
in terms of the generalized Kronecker symbol and the component of the Riemann tensor relative to an orthonormal coframe.
The self-dual anti-self-dual 2-forms are those that diagonalize the Hodge dual or the duality relations defined above in general.
In particular,  left and right duals are identical for basis 2-forms by definition.

The Weyl conformal 2-form is the part of the curvature 2-form that is invariant under the conformal transformations 
of the metric, or equivalently, the coframe scaling defined by the relations
\be\label{ct}
\bar{\theta}^{a}=\sigma\theta^a \Leftrightarrow\bar{g}=\sigma^2g
\ee
with an arbitrary conformal factor $\sigma=\sigma(x^\alpha)$.
With the help of the structure equations (\ref{se1}), one can derive that the connection 1-forms transform as
\be
\bar{\omega}^{ab}
=
\omega^{ab}-(i^ad\ln\sigma)\theta^b+(i^bd\ln\sigma)\theta^a
\ee
under (\ref{ct}), where $d\ln\sigma\equiv\sigma^{-1}d\sigma$. Using this result in connection with the second structure equation (\ref{se2}), one can show that
\be\label{curvature-form-ct}
\bar{\Omega}^{ab}
=
\Omega^{ab}
-
\theta^a\wdg \Sigma^b
+
\theta^b\wdg \Sigma^a
\ee
where the vector valued 1-form $\Sigma^a$ is defined as
\be
\Sigma^a
\equiv
Di^ad\ln\sigma
-
(i^ad\ln\sigma)d\ln\sigma
+
\frac{1}{2}
(i_bd\ln\sigma)(i^bd\ln\sigma)\theta^a
\ee
for convenience.
By defining the antisymmetric tensor-valued 2-form $\Delta^{ab}$ as $\Delta^{ab}\equiv \bar{\Omega}^{ab}-{\Omega}^{ab}$, Eq. (\ref{curvature-form-ct}) can be inverted to have
\be\label{sigma-red}
\Sigma^a
=
-\frac{1}{2}i_b\Delta^{ba}+\frac{1}{12}\theta^ai_ci_b\Delta^{bc}
\ee
by calculating the contractions. As the last step, upon replacing the  result in Eq. (\ref{sigma-red}) back into Eq. (\ref{curvature-form-ct}), one finds
\begin{align}
\bar{\Omega}^{ab}
-
\frac{1}{2}\left(\bar{\theta}^a\wdg \bar{R}^b-\bar{\theta}^b\wdg \bar{R}^a\right)
+
\frac{1}{6}\bar{R}\bar{\theta}^{ab}
=&
{\Omega}^{ab}
-
\frac{1}{2}\left({\theta^a}\wdg {R}^b-{\theta}^b\wdg {R}^a\right)
+
\frac{1}{6}{R}{\theta}^{ab}\label{weyl-ct-def}
\end{align}
by making use of the identity $\theta^a\wdg i_a=\bar{\theta}^a\wdg \bar{i}_a$ that follows readily from the relations in Eq. (\ref{ct}). 

By construction, the expression in Eq. (\ref{weyl-ct-def}) defines the part of the curvature 2-form which remains invariant under the transformation 
in Eq. (\ref{ct}). By denoting  the conformally invariant part of the curvature 2-form as $C^{ab}\equiv\frac{1}{2}C^{ab}_{\fant{ab}cd}\theta^{cd}$, 
the result in Eq. (\ref{weyl-ct-def}) allows one to decompose  the curvature 2-form  into its irreducible parts as 
\be\label{decomposition}
\Omega^{ab}
=
C^{ab}
+
D^{ab}
+
E^{ab}
\ee
where $C^{ab}$ corresponds to the fourth-rank tensor derived above, and $D^{ab}$ can be expressed in terms of the contractions of the curvature 2-form, namely, second-rank trace-free Ricci tensor. The last term denoted by $E^{ab}$ stands for the trace. The second rank part can explicitly be written out in terms of the trace-free 
Ricci 1-form $S^a=R^a-\frac{1}{4}R\theta^a$ as
\be
D^{ab}
\equiv
\frac{1}{2}
\left(
\theta^a\wdg S^{b}
-
\theta^b\wdg S^a
\right),
\ee
whereas the trace part is
\be
E^{ab}
\equiv
\frac{1}{12}R\theta^{ab}.
\ee
The components of the 2-forms defined by the expansions $D^{ab}=\frac{1}{2}D^{ab}_{\fant{ab}cd}\theta^{cd}$ and  $E^{ab}=\frac{1}{2}E^{ab}_{\fant{ab}cd}\theta^{cd}$
are explicitly given by
\begin{align}
D_{abcd}
&=
\frac{1}{2}
(
\eta_{ac}S_{bd}
-
\eta_{bc}S_{ad}
-
\eta_{ad}S_{bc}
+
\eta_{bd}S_{ac}
),
\\
E_{abcd}
&=
\frac{1}{12}R
(
\eta_{ac}\eta_{bd}
-
\eta_{ad}\eta_{bc}
),
\end{align}
respectively.

The definitions (\ref{duality-defs1})-(\ref{duality-defs3}) are valid for the individual irreducible parts of a curvature 2-form defined in (\ref{decomposition}). With respect to duality relations, the irreducible  parts have the properties
\begin{align}
*C^{ab}
&=
+C^{ab}*,
\label{duality-rel1}\\
*D^{ab}
&=
-D^{ab}*,
\label{duality-rel2}\\
*E^{ab}
&=
+E^{ab}*,\label{duality-rel3}
\end{align}
in the  adopted notation and conventions. The derivations of the relations (\ref{duality-rel1})-(\ref{duality-rel2}) are provided in  Appendix below.

The irreducible parts of the curvature 2-form have the following orthogonality relations among themselves:  
\begin{align}
\Omega^{ab}\wdg*C_{ab}
&=
C^{ab}\wdg*C_{ab}
=
C^{ab}*\wdg C_{ab},
\label{curvature-inner-products1}\\
\Omega^{ab}\wdg*D_{ab}
&=
D^{ab}\wdg*D_{ab}
=
-D^{ab}*\wdg D_{ab}
=
S^a\wdg *S_a
=
R_a\wdg *R^a -\frac{1}{4}R^2*1,
\label{curvature-inner-products2}\\
\Omega^{ab}\wdg*E_{ab}
&=
E^{ab}\wdg* E_{ab}
=
E^{ab}*\wdg E_{ab}=\frac{1}{12}R^2*1.\label{curvature-inner-products3}
\end{align}

Consequently, the duality relations for the irreducible parts are combined to have
$
*\Omega^{ab}
=
\Omega^{ab}*+2D^{ab}*
$
or equivalently, rewritten as 
\be\label{dd-decomposition}
\tilde{\Omega}^{ab}
=
-C^{ab}
+
D^{ab}
-
E^{ab}
\ee
in terms of the irreducible parts. As a consequence, one also has the identity $\tilde{\Omega}^{a}_{\fant{a}b}\wdg\theta^b=0$. Moreover,
one can also show that the contraction of double-dual curvature 2-form yields the Einstein 1-form: 
$i_a\tilde{\Omega}^{ab}=\tilde{R}^{ab}_{\fant{ab}ac}\theta^c=G^b=R^b-\frac{1}{2}R\theta^b$, and consequently $i_bi_a\tilde{\Omega}^{ab}=-R$
in accordance with Eq. (\ref{dd-curv-comp}) above.

Four spacetime dimensions are special for quadratic curvature Lagrangians because of the dimensionally continued Euler-Poincar\`e 4-form ($L_{GB}$ Gauss Bonnet topological term). From the above inner products defined in Eqs. (\ref{curvature-inner-products1})- (\ref{curvature-inner-products3}),  and the curvature definitions, one can easily deduce that the square of the Riemann tensor, the square of the left Riemann tensor,  square of the double-dual Riemann tensors  lead to the same expressions in terms of the products of the individual irreducible parts.
The only  exception for the quadratic curvature expressions arises from the inner product between the  curvature of form and double-dual curvature 2-form 
$\Omega_{ab}\wdg *\tilde{\Omega}^{ab}$. On the other hand, this term can be rewritten in a more familiar  form
\be
L_{GB}
\equiv 
-\Omega_{ab}\wdg *\tilde{\Omega}^{ab}
=
\frac{1}{2}\Omega_{ab}\wdg \Omega_{cd}*\theta^{abcd}.
\ee
In terms of  irreducible parts, by making use of  (\ref{decomposition}) and (\ref{dd-decomposition}), and the orthogonality relations above, $L_{GB}$ explicitly reads
\be
L_{GB}
=
C^{ab}\wdg *C_{ab}-R^{a}\wdg *R_a+\frac{1}{3}R^2*1.
\ee
On the other hand, being a topological invariant $L_{GB}$ is an exact four-form, one has the well known result 
\be
C^{ab}\wdg *C_{ab}
=
\left(R^{ab}R_{ab}-\frac{1}{3}R^2\right)*1\qquad \text{(mod d)}
\ee
which is peculiar to four spacetime dimensions.

Finally, it is interesting to note that there is still another topological term expressible in terms of  quadratic curvature components,
namely the gravitational Chern-Simons term with the explicit form
\be\label{cs-term}
L_{CS}
=
\frac{1}{2}\Omega_{ab}\wdg \Omega^{ba}.
\ee
With the help of the first Bianchi identity $\Omega^{a}_{\fant{a}b}\wdg \theta^b=0$ satisfied by  each irreducible curvature parts, it is possible to reduce the Chern-Simons term (\ref{cs-term})  further to the folowing equivalen forms:
\be
L_{CS}
=
\frac{1}{2}C_{ab}\wdg C^{ba}
=
-
\frac{1}{2}C_{ab}\wdg*C^{ba}*
=
\frac{1}{2}C_{ab}*\wdg*C^{ba}.
\ee 

The discussion shows that the suitable quadratic curvature Lagrangian involves the double-dual curvature regarding 
 the constraint that generate a derivative for the Lagrange multiplier by making an essential use of the second Bianchi identity.

\section{The Lanczos' quadratic curvature  Lagrangian and the constraints}\label{L_Lagrangian+Constraints}

In order to construct a tensor potential for the Weyl tensor, it is convenient to  make use of variational principles  
starting from a simple quadratic curvature Lagrangian of the form
\be
L[\tilde{\Omega}^{ab},g]
=
\frac{1}{2}\tilde{\Omega}^{ab}\wdg *\tilde{\Omega}_{ab}
=
-\frac{1}{2}\Omega^{ab}*\wdg *\Omega_{ab}*
=
-\frac{1}{2}\Omega^{ab}**\wdg *\Omega_{ab}
=
-\frac{1}{2}\Omega^{ab}\wdg *\Omega_{ab},
\ee
and subsequently introduce two constraints by the corresponding tensor-valued Lagrange multiplier 1-forms $\lambda_a=\lambda_{ab}\theta^b$  and 
$H_{ab}=H_{abc}\theta^c$ to impose  two identities satisfied by the double-dual curvature tensor. The construction proceeds in parallel to the construction  in \cite{odonnell_spinor_book}  by making use of a coordinate frame.

The double-dual curvature 2-form is assumed to be an independent variable of the Lanzos' Lagrangian  with the metric  tensor playing only a secondary 
role in the construction. Thus, the variation with respect to the double-dual curvature 2-form is given by
\be
\frac{\delta L}{\delta \tilde{\Omega}_{ab}}
=
\frac{\partial L}{\partial\tilde{\Omega}_{ab}}
=
*\tilde{\Omega}^{ab}
\ee
because only the variational derivative with respect to  $\tilde{\Omega}^{ab}$ is of interest. 
At this point the following brief remarks are in order concerning the $p-$form derivatives:
The derivative of a $p$-form with respect to  a $q$ form (assuming $p\geq q$) can be considered as a  linear map
of dimensions $\binom{p}{n} \times\binom{q}{n}$ which is to be equal to  $\binom{p-q}{n}$
consequently, the variational derivative of a volume $n$-form with respect to a $p-$form is well defined 
and  can be expressed in terms of partial derivatives with respect to the corresponding components \cite{kopczynski,dereli-tucker-var}.
Apart from considerations of dimensionality of the derivative mapping,
a $p$-form derivative can also be defined \cite{hehl} by a variational derivative of a volume form for practical purposes.

If one assumes that in place of  the metric,  the coframe basis  1-forms can be  taken as the dynamic variable
 then the variational derivative with respect to $\theta^a$ is
\be
\tilde{\Omega}_{ab}\wdg  \delta *\tilde{\Omega}^{ab}
=
\delta \tilde{\Omega}_{ab}\wdg *\tilde{\Omega}^{ab}
-
\delta\theta^c \wdg
\left(
(i_c\tilde{\Omega}_{ab})\wdg *\tilde{\Omega}^{ab}
-
\tilde{\Omega}_{ab}\wdg i_c*\tilde{\Omega}^{ab}
\right)
\ee
where the $\delta\theta^c$ terms arise from the commutation of the variational derivative $\delta$ with the Hodge dual (or equivalently, the left dual).
It is interesting to note at this point that
\be
(i_cC_{ab})\wdg *C^{ab}
-
C_{ab}\wdg i_c*C^{ab}
=0
\ee
identically, due to the duality relations
\be
C_{ab}\wdg i_c*C^{ab}
=
C_{ab}\wdg i_cC^{ab}*
=
C_{ab}*\wdg i_cC^{ab}
=
i_cC^{ab} \wdg *C_{ab}
\ee
in four spacetime dimensions.

The construction of the Lanczos potential by  use of the variational principles then proceeds then by introducing two (seemingly trivial) constraints. 
One of them  is equivalent to the integrability condition for the curvature 2-form, and the second one that eliminates the second rank and the 
trace parts  in Eq. (\ref{dd-decomposition}). It is convenient to use the Lagrange multipliers technique to impose the constraints on the double-dual curvature 2-form. Subsequently, the constraints are expressed in terms of the double-dual curvature 2-form which helps to eliminate  one of the Lagrange multipliers by using the field equations. Eventually, the remaining one turns out to yield an expression for the  Weyl 2-form. 

It is apparent from the derivation that a similar construction for the Riemann tensor itself would not work \cite{massa,Edgar1987} since the 
Lagrange multiplier is introduced to eliminate the second rank and the scalar trace irreducible parts .
Furthermore, the derivation depends crucially on the duality relations peculiar to the four spacetime dimensions.

The first Lagrange multiplier to be introduced is what is known as the Lanczos' spintensor, which corresponds to antisymmetric tensor-valued 1-forms, 
\be
H_{ab}
=
H_{[ab]}
=
H_{abc}\theta^c,
\ee
and $H_{ab}$ imposes the constraint
\be\label{C1-red}
D*\tilde{\Omega}^{ab}=-\frac{1}{2}\epsilon^{ab}_{\fant{ab}cd}D\Omega^{cd}\equiv 0.
\ee
Thus, constraint (\ref{C1-red}) is equivalent to the second Bianchi identity (the integrability condition for the curvature 2-form) $D\Omega^{ab}=0$. The corresponding constraint 4-form then can be written explicitly in the form
\be
L_{C1}
=
H_{ab}\wdg D*\tilde{\Omega}^{ab}.
\ee
In four spacetime dimensions, without further assumptions regarding the components of the 1-form $H_{ab}$, it has 24 independent  
components. In the construction below, the Weyl tensor is to be expressed in terms of the covariant derivatives  of this third rank tensor 
$H_{abc}$ and since the Weyl tensor has 10 algebraically independent components one introduces further restrictions on the components
of $H_{abc}$ to reduce the number of its independent components to 10, effectively. It is convenient to return to the issue of constraining
the components of $H_{ab}$ after reducing the Lanczos' Lagrangian below.

A second Lagrange multiplier, a symmetric tensor-valued 0-form $\lambda_{ab}=\lambda_{(ab)}$, is introduced to impose the constraint $G_{ab}=0$
(Einstein field equations for vacuum) with a corresponding  constraint 4-form
\be
L_{C2}
=
\lambda_{ab}G^{ab}*1.
\ee

In order to put the $L_{C2}$ into a suitable form expressed in terms of double-dual curvature 2-form, one starts with rewriting the  
the constraint as an inner product of two tensor-valued forms as
\be
\lambda_{ab}G^{ab}*1
=
\lambda_{a}\wdg *G^a
\ee
where $\lambda_{a}\equiv\lambda_{ab}\theta^b$ and $G^{a}\equiv G^{a}_{\fant{a}b}\theta^b$ stands for  the Einstein 1-form.
Furthermore, let us recall that the Einstein form can be expressed as the product of the curvature 2-forms as
\be
*G^a
=
-\frac{1}{2}\Omega_{ab}\wdg *\theta^{abc}
=
\theta_b\wdg*\tilde{\Omega}^{ba}.
\ee
Consequently, the constraint 4-form $L_{C2}$ can be rewritten in the form
\be
L_{C2}
=
\frac{1}{2}\left(
\theta^a\wdg \lambda^b
-
\theta^b\wdg \lambda^a
\right)
\wdg *\tilde{\Omega}_{ba}
\ee
Recently, O'Donnell \cite{ODonnell-2011} attributed a physical significance to the scalar trace $\lambda\equiv i_a\lambda^a$
of the Lagrange multiplier forms $\lambda_a=\lambda_{ab}\theta^b$. In a more general context, Roberts \cite{roberts-interpret} discussed the physical significance of 
the Lanczos potential in parallel to the Bohm-Aharonov affect for a $U(1)$ gauge potential.

Returning back to the construction, eventually, the extended Lanczos' Lagrangian 4-form takes the form
\be
L'
=
L+L_{C1}+L_{C2}
\ee
and that it  depends on  three independent variables, all of which are expressed as tensor-valued forms:
\be
L'
=
L'[\tilde{\Omega}^{a}_{\fant{a}b}, H_{ab}, \lambda_{a}].
\ee

The dependence of the variables on a Riemannian  metric $g$ enters into the discussion through the following assumptions on the geometrical variables: 
\begin{itemize}
\item[(1)]
The curvature 2-form is derived from the Levi-Civita connection 1-form $\omega_{ab}=\omega_{[ab]}$, and therefore one has $\Omega_{ab}=\Omega_{[ab]}$ and consequently $H_{ab}=H_{[ab]}$, 
\item[(2)] The inner product for the tensor valued $p$-forms  provided by the left (Hodge) dual can be defined only after a Riemannian metric is introduced.  
\end{itemize}

All of the tensorial quantities in the discussion are assumed to belong to the (pseudo-)Riemannian geometry. Furthermore, since the metric structure plays a secondary role in the variational procedure, the dependence of the metric tensor is suppressed.

Explicitly, the extended Lagrangian
\be
L'
=
\frac{1}{2}\tilde{\Omega}^{ab}\wdg *\tilde{\Omega}_{ab}
+
H_{ab}\wdg D*\tilde{\Omega}^{ab}
+
\lambda_{ab}G^{ab}*1
\ee
can be written in the form
\be\label{lag_form2}
L'
=
\left[
\frac{1}{2}\tilde{\Omega}^{ab}
+
DH^{ab}
+
\frac{1}{2}\left(\theta^a\wdg\lambda^b-\theta^b\wdg\lambda^a \right)\right]
\wdg *\tilde{\Omega}_{ab}
-
d\left(
H_{ab}\wdg *\tilde{\Omega}^{ab}
\right)
\ee
which renders the variational derivative of the extended Lagrangian with 
respect to the double-dual curvature 2-form almost trivial. More precisely,
$\delta L'/\delta \tilde{\Omega}_{ab}=0$ is equivalent to the vanishing of 
the derivative with respect to double-dual 2-form and it readily yields
\be
\tilde{\Omega}^{ab}
=
-
DH^{ab}
-
\frac{1}{2}\left(\theta^a\wdg\lambda^b-\theta^b\wdg\lambda^a \right)
\ee
The field equation now can be used to express the Lagrange multiplier $\lambda
_a$ in terms of the remaining variables.
By calculating two successive  contractions, one obtains 
\be
\lambda^a
=
-
R^a
+
\frac{1}{3}R\theta^a
-
i_bDH^{ba}
-
\frac{1}{6}
\theta^ai_bi_cDH^{bc}.
\ee 
Evidently, from the variational derivative with respect to the Lagrange multiplier $\lambda_a$, namely,
\be
\frac{\delta L'}{\delta \lambda_b}
=
\theta_a\wdg *\tilde{\Omega}^{ab}=0,
\ee
one re-obtains the constraint $G_{ab}=0$, or equivalently, $D^{ab}=0$ and $E^{ab}=0$. In addition, 
by making use of the decomposition (\ref{decomposition}), the constraint arising from $L_{C1}$ entails the equality $\tilde{\Omega}^{ab}=C^{ab}$. 
Replacing these results back into the original Lagrangian (\ref{lag_form2}), 
one obtains a reduced Lagrangian  of the form
\be\label{reduced_lag2}
L'_{red.}
=
\left[
-
\frac{1}{2}C^{ab}
+
DH^{ab}
-
\frac{1}{2}\theta^a\wdg i_c DH^{cb}
+
\frac{1}{2}\theta^a\wdg i_c DH^{cb}
-
\frac{1}{6}\theta^{ab}\wdg i_ci_d DH^{cd}
\right]\wdg *C_{ab}
\ee
depending on the reduced variables $C^{ab}$ and $H^{ab}$.

The constraint equation $\delta L'/\delta \lambda_a=0$  can be used in the variational derivative with respect to the double-dual curvature 2-form to have 
\be
\frac{\delta L'}{\delta\tilde{\Omega}^{ab}}\Biggr\rvert_{D_{ab}=E_{ab}=0}
=
\frac{\partial L'}{\partial\tilde{\Omega}^{ab}}\Biggr\rvert_{D_{ab}=E_{ab}=0}
=
\frac{\delta L'_{red.}}{\delta C_{ab}}.
\ee
Consequently, the reduced field equations  can be rewritten in the form
\be
*C^{ab}
=
\frac{\partial L'_{red.}}{\partial C_{ab}}.
\ee
Finally, one can read off the above variational derivative  from Eq. (\ref{reduced_lag2}) that can  be explicitly  written as
\be\label{lc-form1}
C^{ab}
=
DH^{ab}
-
\frac{1}{2}\theta^{a}\wdg i_cDH^{cb}
+
\frac{1}{2}\theta^{b}\wdg i_cDH^{ca}
-
\frac{1}{6}\theta^{ab}\, i_ci_dDH^{cd},
\ee 
as an equation for  the remaining 2-form variables, namely, the Weyl 2-form $C^{ab}$  and $DH^{ab}$ up to an overall factor which can be absorbed into the potential by a redefinition. 

One can show that the compact expression on the right-hand side of (\ref{lc-form1}) 
can be used to derive  the corresponding coordinate expression. As is discussed  below, the Lanczos Lagrangian extended by the Lagrange multipliers has  special \emph{gauge invariances} related to the Lagrange multipliers  that lead to further simplification of the main result in Eq. (\ref{lc-form1}) above. 

Before deriving the components of the Weyl  tensor from (\ref{lc-form1}) to facilitate  a comparison with more familiar tensor component expressions, it is convenient to simplify (\ref{lc-form1}) further.

It is straightforward to verify that the expression on the right-hand side is trace-free, as expected on consistency grounds
since one has $i_a C^{ab}=0$ by definition. On the other hand, let us recall that the first Bianchi identity $\Omega^{a}_{\fant{a}b}\wdg \theta^b=0$ is satisfied 
by individual parts of the curvature, $C^{ab}$, $D^{ab}$ and $E^{ab}$, separately and therefore the expression on the right hand side 
is to satisfy the Bianchi identity, too. 
Unlike the trace-free property, the first Bianchi identity $C^{a}_{\fant{a}b}\wdg \theta^b=0$ is not satisfied identically, and therefore introduces  restrictions on the Lagrange multiplier $H_{ab}$. Thus, the first Bianchi identity for the Weyl 2-form leads to a constraint on both $H_{ab}$, and consequently on $DH_{ab}$ as well. Explicitly, by multiplying (\ref{main-result}) with $\theta_{ab}$ and using the first Bianchi identity for the Weyl 2-form one ends up with a scalar identity
of the form
\be
DH_{ab}\wdg \theta^{ab}=0,
\ee
which can be rewritten in a more conveniently as
\be\label{l-pot-bi}
D(H_{abc}\wdg \theta^{abc})=0.
\ee
Finally, note that the differential identity (\ref{l-pot-bi}) is identically satisfied  if $H_{abc}$ satisfies the algebraic relation $H_{(abc)}=0$, as indicated by 
the expression inside the brackets. 
Thus, the constraint in Eq. (\ref{l-pot-bi}) is not a gauge condition and  is closely related to the first Bianchi identity  satisfied by the Riemann tensor.  
Eqs. (\ref{l-pot-bi}) amounts to six relations among  the independent components of $H_{abc}$. 

The use of the exterior algebra together with the covariant exterior derivative is justified by the compact expression
(\ref{main-result}). For instance, the expression $DH_{ab}$ can be rewritten in an expanded form as
\be\label{exp-DH}
DH_{ab}
=
\frac{1}{2} (i_{c}DH_{abd}-i_{d}DH_{abc})\theta^{cd}
\ee
where the expression on the right-hand side is  antisymmetric with respect to $c\leftrightarrow d$ which follows from the properties of the covariant exrerior 
derivative and $D\theta^{a}=0$. Furthermore, one can show that the symmetry of the indexes in (\ref{exp-DH}) under the interchange $ab\leftrightarrow cd$ follows the relation $H_{(abc)}=0$. Thus, as  a consequence of (\ref{l-pot-bi}), the components of the antisymmetric tensor-valued 2-form $DH_{ab}$ has the same index symmetries with respect to first and second pair of indexes as the Riemann and Weyl tensors.

In the gauge theory, for example, in the case of the $U(1)$ gauge theory, the potential $A$ is not unique and either $A$ or $A+d\varphi$  locally generates the same field strength $F=dA$, as a consequence of $dd\equiv 0$. There is a similar gauge arbitrariness in the definition of the Lanczos' tensor potential $H_{ab}$ to be discussed below. Thus, by making use of the gauge freedom it is possible to choose a potential $H_{abc}$ satisfying the  conditions
\be\label{restrictions_2-3}
H_{ab}^{\fant{ab}b}=0 \quad \mbox{and} \quad i^{c}DH_{abc}=0.
\ee
The gauge conditions in Eq. (\ref{restrictions_2-3}) are usually referred to as trace-free (algebraic) and divergence-free (differential) gauge conditions, respectively.

Assuming that the potential $H_{ab}$ also satisfies Eqs. (\ref{restrictions_2-3}), the expression (\ref{lc-form1}) further reduces to the form
\be\label{main-result}
C^{ab}
=
DH^{ab}
-
\frac{1}{2}\left(
\theta^{a}\wdg i_cDH^{cb}
-
\theta^{b}\wdg i_cDH^{ca}
\right)
\ee 
This expression, known as Weyl-Lanczos equations, implies the peculiar fact that the fourth rank Weyl 2-form can be generated locally by covariant  derivatives of a lower rank tensor $H_{abc}$ along with additional algebraic symmetry/anti-symmetry relations among its indexes. 

 As a result, with the additional relations satisfied by the Lanczos potential, explicitly the relations,
\be\label{index-relations}
H_{ba}^{\fant{aa}a}=0, \qquad H_{(abc)}=0, \qquad i^cDH_{abc}=0,
\ee
the number of algebraically  independent components of the Lanczos potential reduce from 24 to 10, effectively.
As remarked above, the constraints on the Lanczos potential  $H_{ab}$ can also be interpreted as a kind of gauge fixing procedure to be discussed below.

In the exterior algebra notation, relations (\ref{index-relations}) can be expressed in the form
\be
i^aH_{ab}=0,\qquad H_{ab}\wdg \theta^{ab},\qquad D*H_{ab}=0, 
\ee
respectively.

\section{A comparison of the Lanczos potential expressions}\label{comparison}

It is possible to verify  expression (\ref{main-result}) by using it to derive the corresponding coordinate expression for the Weyl tensor $C_{abcd}$.

First let us note that, by taking the expansion of the Weyl 2-form into its components 
\be
C_{ab}
=
\frac{1}{2}C_{abcd}\theta^{cd}
\ee
into account, one can obtain the components of the Weyl tensor with respect to an orthonormal coframe explicitly in terms of $H_{ab}$.

One can show that the first term on the right-hand side of (\ref{main-result}) can be written in the convenient form
\be\label{p1}
DH_{ab}
=
-\frac{1}{2}
\left(
i_d DH_{abc}
-
i_cDH_{abd}
\right)
\theta^{cd}
\ee
whereas the second term on the right-hand side can be written in the form
\be\label{p2}
-\frac{1}{2}
\left(
\theta_a\wdg i_c DH^{c}_{\fant{c}b}
-
\theta_b\wdg i_c DH^{c}_{\fant{c}a}
\right)
=
-\frac{1}{4}
\left(
H_{bd}\eta_{ac}
-
H_{bc}\eta_{ad}
-
H_{ad}\eta_{bc}
+
H_{ac}\eta_{bd}
\right)
\ee
where the abbreviation $H_{ab}\equiv i_cDH^{c}_{\fant{c}ab}$ is used for convenience of notation.

The symmetry of the pair of indexes $(ab)$ and $(cd)$ for $C_{abcd}$ under the interchange $(ab)\leftrightarrow(cd)$ is a consequence of the 
the Bianchi identity. On the other hand, it is noted in the above  discussion that the expression on the right-hand side in Eq. (\ref{main-result}) 
does not satisfy it identically. The expression in Eq. (\ref{main-result}) can be made to satisfy the Bianchi identity by assuming that $H_{(abc)}=0$ is valid, and consequently, the symmetrization of  expressions (\ref{p1}) and (\ref{p2}) is allowed. Finally, with the replacement $i_aDh_{bcd}\equiv\nabla_{a}H_{bcd}$, one ends up with
\begin{align}
C_{abcd}
=&
-\frac{1}{2}
\big(
\nabla_{c}H_{abd}-\nabla_{d}H_{abc}+\nabla_{a}H_{cdb}-\nabla_{b}H_{cda}
\nonumber\\
&
+
H_{(bd)}\eta_{ac}-H_{(bc)}\eta_{ad}-H_{(ad)}\eta_{bc}-H_{(ac)}\eta_{bd}
\big)\label{castillo-expresion}
\end{align}
and this is the expression given,  for example, by Castillo \cite{castillo} who also discusses the  Lanczos potential for Petrov type D spacetimes, up to an overall multiplicative factor of $-\frac{1}{2}$. The expression is usually referred to as Weyl-Lanczos relation in the literature.

The coordinate expression (\ref{castillo-expresion}) can readily be derived with the help of the expression in Eq. (\ref{main-result}) obtained by  using the exterior algebra of differential forms. On the other hand,  it  is slightly more difficult  to obtain an expression of the form  given  in Eq. (\ref{main-result})  from the expression in Eq. (\ref{castillo-expresion}) defined with respect to a coordinate basis.  
 
Although it is often claimed that the Lanczos potential is a more fundamental geometrical object than the Weyl tensor itself, it received not much attention that would justify the claim, perhaps, because of the fact that it is fairly difficult to solve for a corresponding Lanczos' tensor potential by making use of the expression in  either (\ref{main-result}) or (\ref{castillo-expresion}) for a given Weyl tensor.

\section{Gauge invariance for the extended  Lanczos Lagrangian}\label{gauge_invariance}

As mentioned in the previous sections, there is  gauge freedom in the choice of the Lanczos tensor potential.
The invariance of the Lanczos' (extended) Lagrangian under transformation of the Lagrange multipliers was discussed first by Takeno \cite{takeno}, later
the gauge invariance was also discussed by Anderson and Edgar \cite{andersson} using  2-spinors.
For some time, the fact that the Lanzcos spintensor generates the Weyl tensor remained unnoticed before Takeno \cite{takeno} explicitly showed that Lanzcos spintensor $H_{ab}$ locally generates the Weyl tensor.

It is always possible to chose a Lanczos potential such that it is (1) trace-free and (2) divergence-free
by making use of gauge freedom that can be introduced at the level of the Lagrangian. 
More precisely, it is possible to verify in a straightforward manner that the Lanczos' (extended) Lagrangian 
remains invariant, that is
\be
L'[\tilde{\Omega}^{a}_{\fant{a}b}, \lambda^a, H_{ab}]
=
L'[\tilde{\Omega}^{a}_{\fant{a}b}, \lambda'^a, H'_{ab}],
\ee
under the transformations of the Lagrange multiplier defined by
\begin{align}
&H_{ab}
\mapsto H'_{ab}=H_{ab}+V_b\theta_a-V_a\theta_b,
\label{gt1}\\
&
\lambda^a\mapsto
\lambda'^a=\lambda^a+2DV^a.\label{gt2}
\end{align}
in terms an arbitrary vector field $V_a$.

Using the above gauge invariance, one can have an potential $H_{abc}$ such that $i^b H_{ab}=H_{ab}^{\fant{ab}b}=0$.
In component form, the gauge transformations (\ref{gt1}) read
\be
H'_{abc}
=
H_{abc}
+
\eta_{ac}V_{b}
-
\eta_{bc}V_a.
\ee
Assuming that $H_{ab}^{\fant{ab}b}\neq 0$, by contraction of the above equation with $\eta^{bc}$ one finds
\be
H_{ab}^{'\fant{ab}b}
=
H_{ab}^{\fant{ab}b}
-
3V_a.
\ee
Thus, one has the freedom to choose $V_a=\frac{1}{3}H^{\fant{ab}b}_{ab}$ so that $H^{'\fant{ab}b}_{ab}=0$ in the new gauge.

The gauge freedom can further be used to make $H_{ab}$ a divergence-free tensor-valued 1-form as well. Note that
divergence free condition can be expressed in the form $D*H_{ab}=0$ as well since one has 
\be
D*H_{ab}
=
i_cDH_{ab}^{\fant{ab}c}*1.
\ee
By using the Hodge dual of the gauge  transformations (\ref{gt1}), one has
\be\label{gauge-fix-2}
D*H'_{ab}
=
D*H_{ab}
+
DV^b\wdg *\theta^a
-
DV^a\wdg *\theta^b
\ee
and therefore even if one has $D*H_{ab}\neq0$, one can set $D*H'_{ab}=0$ by requiring that the right-hand side of Eq. \ref{gauge-fix-2} vanishes.
The constraints discussed above can be introduced independently. 

Finally, note that the divergence-free gauge condition can be combined with the algebraic relation $DH_{(abc)}=0$ (Bianchi identity) to show that the relation of the form
\be\label{additional-cov-id}
i^cDH_{acb}
=
i^cDH_{bca}
\ee 
is valid for the covariant derivatives of $H_{abc}$ as well. The identity in Eq. (\ref{additional-cov-id}) can be regarded as  an identity similar to the 
contracted fist Bianchi identity for the Weyl tensor.

\section{A Lanczos potential for a Petrov type N gravitational wave}\label{Lanczos-pot-for-pp-waves}

In the formulation of the Lanczos potential above, an orthonormal coframe is adopted; however all the formulae are valid for any rigid coframe with constant metric components. As an  application of expression (\ref{main-result}) obtained above, one can calculate the Lanczos potential for the 
$pp$-wave metric using a complex null coframe as follows.

It is well known that the $pp$-wave the curvature tensor, describing a  plane-fronted gravitational wave with parallel rays,
is linear in the metric components, and therefore the Einstein field equations are linear in the derivatives of the metric components. 
In terms of a set of complex null coordinates $\{x^\mu\}=\{u, v, \zeta, \bar{\zeta}\}$ for $\mu=0, 1, 2, 3$, the $pp$-wave metric can be written in the form
\be\label{pp-metric}
g=
-du\ot dv-dv\ot du-2H(u,\zeta,\bar{\zeta})du\ot du+d\zeta\ot d\bar{\zeta}+d\bar{\zeta}\ot d\zeta
\ee
where $H=H(u,\zeta,\bar{\zeta})$ is the profile function for the gravitational waves propagating  in the $\partial_v$ direction.
In this section a bar over the quantity denotes complex conjugation. One can introduce a Newman-Penrose (NP) \cite{NP} complex null coframe 
\be\label{pp-coframe-def}
\theta^0=k=du,\quad \theta^1=l=dv+Hdu,\quad \theta^2=m=d\zeta,\quad \theta^3=\bar{m}=d\bar{\zeta}.
\ee
The only nonvanishing exterior derivative of basis coframe 1-forms is explicitly given by
\be
dl
=
-H_{,\zeta}k\wdg m-H_{,\bar{\zeta}}k\wdg \bar{m},
\ee 
and consequently, the Cartan's first structure equations (\ref{se1}) yield the only nonvanishing Levi-Civita connection
\be
\omega^{1}_{\fant{1}2}
=
H_{,\zeta}du
\ee
with respect to the null coframe (\ref{pp-coframe-def}) and $H_{,\zeta}\equiv\partial_{\zeta}H$. The nonvanishing components of the corresponding curvature 2-form, which is linear in the profile function, takes the form
\be\label{pp-curvature}
\Omega^{1}_{\fant{a}2}
=
d\omega^{1}_{\fant{1}2}
=
-
H_{,\zeta\zeta}k\wdg m
-
H_{,\zeta\bar{\zeta}}k\wdg \bar{m}
\ee
where the numerical indexes refer to the null coframe.
The self-dual part of the curvature corresponding to the $k\wdg m$ component corresponds to the only nonvanishing 
Weyl tensor component \cite{Baldwin-Jeffrey,baykal-pp,baykal-pp2}. For vacuum, the Einstein field equations are satisfied if $H_{\zeta\bar{\zeta}}=0$, so that
\be
\Omega^{1}_{\fant{a}2}
=
C^{1}_{\fant{a}2}
=
H_{,\zeta\zeta}k\wdg m
\ee
in this case.
Therefore, one can see that the Lanczos potential has only one complex component
\be\label{pp-l-pot}
H^{1}_{\fant{a}2}
=
H_{,\zeta}du
\ee
with respect to the null coframe (\ref{pp-coframe-def}).  The potential (\ref{pp-l-pot}) satisfies the trace-free and divergence-free
gauge conditions and the Bianchi identity $H_{(abc)}=0$, identically. Moreover, for the $pp$-wave metric (\ref{pp-metric}), one has 
$DH^{1}_{\fant{a}2}=d H^{1}_{\fant{a}2}$, and consequently one can show that (\ref{main-result}) reduces to
\be\label{pp-weyl-2-form}
C^{1}_{\fant{a}2}
=
dH^{1}_{\fant{a}2}
\ee
in this case. Finally, by comparing (\ref{pp-weyl-2-form}) with the curvature expression (\ref{pp-curvature}), one ends up with the result  
\be\label{pp-pot}
H_{020}
=
-H_{\zeta}.
\ee 
The nonvanishing component of the Lanczos potential (\ref{pp-pot}) can also be expressed as the contraction $l^{\alpha}\bar{m}^\beta l^\mu H_{\alpha\beta\mu}$ corresponding to $H_7$ in \cite{odonnell_spinor_book} up to the constant multiple arising from our definition of $H_{abc}$. The only nonvanishing component of the Lanczos potential for the complex Weyl scalar $\Psi_4$ turns out to be expressed in terms of the spin coefficient $\nu$ (and its complex conjugate) 
\cite{odonnell_spinor_book,bonilla} for the $pp$-wave  metric (\ref{pp-metric}).

It is interesting to note that previously Bergqvist \cite{bergqvist}  obtained a Lanczos potential for the mathematically more involved Kerr spacetime in terms of a flat connection using the Geroch-Held-Penrose formalism  also calculating the total energy using the Lanczos potential he constructed.

\section{Is the Riemann tensor derivable from a tensor potential?}\label{massa-sec}

The very important question  whether an acceptable solution to the Weyl-Lanczos relations exists is discussed by a number of authors focusing
on different, yet equally interesting aspects. For instance, Bampi  and Caviglia \cite{Bampi1983,Bampi1984,edgar-hoglund1997} proved the local existence theorem for solutions of the Weyl-Lanczos equations in four dimensions. Later, Andersson and Edgar \cite{andersson} presented a more concise proof of their existence for the Lanczos' tensor  potential. 

In $n$ spacetime dimensions, the total number of algebraically independent components  of a tensor (denoted by $N$) with  index symmetries/antisymmetries
$C_{abcd}=C_{[ab]cd}=C_{ab[cd]}$ and $C_{(abc)d}=0$ is given by the formula
\be\label{n-formula}
N
=
\frac{1}{12}n(n+1)(n+2)(n-3).
\ee
Thus, for $n=3$,  the number of dimensions is not sufficiently many to support the  definition of a fourth rank tensor with   index symmetries and antisymmetries which is the same as the Weyl curvature tensor. In this case, the conformal invariance is encoded in another important tensor, namely the Cotton tensor \cite{cotton}. But interestingly, there is still a tensor potential for a Riemann tensor in three dimensions  \cite{chrusciel,dolan-gerber}. On the other hand, recently in Ref. 
\cite{Edgar2004}, a new (2,3) double-form  type tensor potential of the form $P^{ab}_{\fant{ab}cde}$ is introduced for the Weyl tensor in  
$n\geq4$ dimensions.

The existence of a tensor potential for the full Riemann tensor in four dimensions is also discussed in the literature.
The heading to this section is the title of the  paper \cite{massa} by Massa and Pagani who answered the question in the negative in a technically elegant manner. 
They proved that it is not possible to derive a Riemann tensor from a third rank tensor by using  tensor-valued forms. Among other things, the discussion in 
\cite{massa} also suggests that the use of exterior algebra has a computational advantage over tensorial methods. (Cf.  the discussion in Ref. \cite{Edgar1987} using tensorial methods and components.) Such a technical  overlap makes it convenient to present the discussion by Massa and Pagani in the notation introduced above. To put the discussion in \cite{massa} briefly, the authors investigated the integrability conditions for the defining relation for the tensor potential. 
They assumed that such a potential exists in general, and then by calculating successive covariant exterior derivatives, they showed that one eventually arrives at a constraint on the (pseudo-)Riemannian geometry that is not assumed to hold in general.

Disregarding the construction by a variational procedure, they started their discussion with the assumption that $H_{ab}=H_{abc}\theta^c$ generates a fourth rank tensor $W_{abcd}$ with the  antisymmetries 
\be\label{massa-anti-sym}
W_{abcd}=W_{[ab]cd}=W_{ab[cd]}
\ee
which can be expressed in the form
\be\label{massa-decomp}
W^{ab}
=
\Omega^{ab}
+
Q^{c}\wdg *\theta^{ab}_{\fant{ab}c}
\ee
where $Q^a\equiv Q^{a}_{\fant{a}b}\theta^b$ is a vector-valued 1-form.
The  decomposition (\ref{massa-decomp}) is valid since  a fourth-rank tensor satisfying (\ref{massa-anti-sym}) with the additional antisymmetry with respect to the first and the second pair of indexes $W_{abcd}=-W_{cdab}$ can uniquely be represented by a second rank tensor $Q^{a}_{\fant{a}b}$ with $Q^{a}_{\fant{a}a}=0$.
As a consequence of this bijective relation, and also assuming that  the Riemann tensor  can be adopted for the symmetric bivector-tensor part,  one ends up with the decomposition in Eq. (\ref{massa-decomp}). It is crucial to note that the decomposition in Eq. (\ref{massa-decomp}) is not valid only for a particular class of Riemannian geometry initially, and it is assumed to hold in general. 

The second step is to introduce a general potential from which  the 2-form $W^{ab}\equiv\frac{1}{2}W^{ab}_{\fant{ab}cd}\theta^{cd}$  
can be locally generated from $H_{ab}$ as
\be\label{massa-der-def}
DH^{ab}
=
\Omega^{ab}+Q^{c}\wdg *\theta^{ab}_{\fant{ab}c}.
\ee

The consistency of the definition in Eq. (\ref{massa-der-def}) in connection with the identity $R_{(abc)d}=0$ requires  that the symmetry conditions $H_{(abc)}=0$ and $Q^a\wdg *\theta_a=Q^{a}_{\fant{c}a}*1=0$ are satisfied. These additional conditions can be obtained by wedging (\ref{massa-der-def}) with $\theta_{ab}$ and then using the first Bianchi identity for the Riemannian curvature 2-form.

The integrability condition for the definition (\ref{massa-der-def}) can  be simply found by calculating the covariant exterior derivative and it turns out to be
\be\label{massa-first-der}
\Omega^{a}_{\fant{a}c}\wdg H^{cb}
-
\Omega^{b}_{\fant{a}c}\wdg H^{ca}
=
DQ^{c}\wdg *\theta^{ab}_{\fant{ab}c}
\ee
by making use of the second  Bianchi identity satisfied by the curvature 2-form.
In turn, the integrability condition for the resultant equation in (\ref{massa-first-der}) can simply be found by calculating its covariant exterior derivative.
One finds
\be\label{massa-second-der}
\Omega^{a}_{\fant{a}c}\wdg DH^{cb}
-
\Omega^{b}_{\fant{a}c}\wdg DH^{ca}
=
DDQ^{c}\wdg *\theta^{ab}_{\fant{ab}c}.
\ee
Finally, by using  the original definition in Eq. (\ref{massa-der-def}) back in  result (\ref{massa-second-der}) and $DDQ^a=\Omega^{a}_{\fant{a}b}\wdg Q^b$,
and subsequently rearranging the resulting terms one ends up with the identity 
\be\label{massa-last}
Q^c\wdg DD*\theta^{ab}_{\fant{ab}c}=0
\ee 
which vanishes identically as a consequence of the first Bianchi identity $DD\theta^a=\Omega^{a}_{\fant{a}b}\wdg \theta^b=0$ without imposing a restriction either on $\Omega^{ab}$ or  $Q^a$. Thus, the procedure of taking successive covariant exterior derivatives of the definition in Eq. (\ref{massa-der-def}) terminates in the second step. 

Next, let us consider the integrability conditions of the Eq. (\ref{massa-first-der}) with symmetrized free indexes. For this purpose, one first considers an equation for 3-forms of the form
\be\label{3-form-eqn}
P^{ab}
=
\theta^a\wdg Z^b-\theta^b\wdg Z^a,
\ee
for $a,b=0, 1, 2, 3$. By definition,  $Z^a=\frac{1}{2}Z^{a}_{\fant{a}bc}\theta^{bc}$ is a vector valued 2-form which renders  
$P^{ab}=P^{[ab]}=\frac{1}{6}P^{ab}_{\fant{ab}cde}\theta^{cde}$  an antisymmetric tensor-valued 3-form. When it is written in component form
Eq. (\ref{3-form-eqn})  establishes a bijective relation between the components of the tensor valued-forms $Z^a$ and $P^{ab}$. By calculating the contractions of
Eq. (\ref{3-form-eqn}), it can be inverted to express $Z^a$ in terms of $P^{ab}$ as an equation for 2-forms in the form
\be\label{2-form-eqn}
Z^a
=
i_bP^{ba}
+
\frac{1}{4}\theta^a \wdg i_{b}i_{c}P^{bc}.
\ee
Consequently, the relation in Eq. (\ref{2-form-eqn}) can be used to derive  the symmetric version of the general relation in Eq. (\ref{3-form-eqn}) as
\be\label{sym-id}
i_{c}\left(
\theta^a\wdg P^{bc}+\theta^b\wdg P^{ac}
\right)
=
\theta^a\wdg Z^b+\theta^b\wdg Z^a.
\ee

Returning to the integrability conditions given in Eq. (\ref{massa-first-der}),  one can readily rewrite it as an equation for 3-forms  in the form 
\be\label{hodge-dual-massa-first-der}
\theta^a\wdg DQ^b-\theta^b\wdg DQ^a
=
\epsilon^{ab}_{\fant{ab}cd}
\Omega^{c}_{e}\wdg H^{ed}
\ee
by using the expression for $*\theta^{ab}_{\fant{ab}c}=\epsilon^{ab}_{\fant{ab}cd}\theta^{d}$ on the right-hand side in Eq. (\ref{massa-first-der}), and subsequently taking the permutation symbol to the left-hand side. Therefore, the integrability condition for the derived relation in Eq. (\ref{hodge-dual-massa-first-der}) is equivalent to that of (\ref{massa-first-der}). Evidently, Eq. (\ref{hodge-dual-massa-first-der}) has the general form given in Eq. (\ref{3-form-eqn}) with the identifications $Z^a=DQ^a$ and $P^{ab}=\epsilon^{ab}_{\fant{ab}cd} \Omega^{c}_{\fant{c}e}\wdg H^{ed}$. Accordingly, the identifications allow one to rewrite 
Eq. (\ref{sym-id}) conveniently as
\be\label{sym-second-step1}
\theta^a\wdg DQ^b+\theta^b\wdg DQ^a
=
i_{c}\left(
\theta^a \wdg \epsilon^{bc}_{\fant{ab}mn} \Omega^{m}_{\fant{c}d}\wdg H^{dn}
+
\theta^b \wdg \epsilon^{ac}_{\fant{ab}mn} \Omega^{m}_{\fant{c}d}\wdg H^{dn}
\right).
\ee
Eq. (\ref{sym-second-step1}) is the symmetrical version of the integrability condition in Eq. (\ref{massa-first-der}). The integrability condition for the derived relation in Eq. (\ref{sym-second-step1}) then takes the form 
\be\label{non-int}
\left(
\theta^a\wdg \Omega^{b}_{\fant{a}c}+\theta^b\wdg \Omega^{a}_{\fant{a}c}
\right)
\wdg Q^c
=
-Di^{c}\left[
\left(
\theta^a \epsilon^{b}_{\fant{a}cmn} 
+
\theta^b \epsilon^{a}_{\fant{a}cmn} \right)\wdg \Omega^{m}_{\fant{c}d}\wdg H^{dn}
\right].
\ee
As a result, the integrability condition for the symmetrical part of the expression derived in the first step, namely,
$
\theta^a\wdg DQ^b
=-D(\theta^{(a}\wdg Q^{b)}+\theta^{[a}\wdg Q^{b]})
$
 does not vanish identically as a consequence of the previous equations. 

In particular, the traces of the terms on the left-hand side of Eq. (\ref{non-int}) vanish identically as a consequence of the first Bianchi identity  
$\Omega^{a}_{\fant{a}b}\wdg \theta^b=0$.  Thus,  by calculating the trace of the right-hand side, the trace of Eq. (\ref{non-int}) then boils down to 
\be\label{disproof}
Di_{a}
\left(
\Omega_{bd}\wdg H^{d}_{\fant{a}c}
\wdg *\theta^{abc}
\right)
=
\theta^{ab}
\wdg
D*\left(
\Omega_{ac}\wdg H^{c}_{\fant{a}b}
\right)
=
0.
\ee

Eventually, as  was first noted in ref. \cite{massa}, the trace of the constraint  allows one to conclude that Eq. (\ref{non-int}) is bound to generate  additional constraints on the Riemann curvature tensor.  More explicitly, by  use of the original definition in Eq. (\ref{massa-der-def}), and the first Bianchi identity, one can see that Eq. (\ref{disproof}) can  be expressed concisely in the following form:
\be\label{disproof2}
\frac{1}{2}
\left(
\Omega_{ab}
+
\epsilon_{abcd} Q^c \wdg \theta^{d}
\right)
\wdg  *\left[\Omega^{ab}+ \frac{1}{2}\left(\theta^{a} \wdg  R^{b}-\theta^{b} \wdg  R^{a}\right)\right]
+
H^{d}_{\fant{a}ca}\theta^{bc} \wdg D*\left(\Omega_{bd}\wdg \theta^a\right)
=
0.
\ee
In terms of the tensorial components, the constraint in Eq. (\ref{disproof2}) can symbolically be rewritten in the form
\be\label{final-der-symbolic}
R^2
+
QR
+
H\nabla R
=
0
\ee
where $\nabla$ stands for the covariant derivative, and $Q, R$ and $H$ stand for the components $Q^{a}_{\fant{a}b}, R_{abcd}$ and $H_{abc}$, respectively.

More importantly, the covariant derivative of the trace in Eq. (\ref{final-der-symbolic}) has the potential to generate further constraints in the general case. The consequences of the constraint given in Eq. (\ref{non-int}) and its trace in Eq. (\ref{disproof2}) have been discussed previously in \cite{Edgar2003} in detail in a broader context  including the Riemann-Lanczos problem.  Although the argumentation above  does not exclude the possibility that such a potential exists under special circumstances \cite{Bampi1983,Bampi1984,Edgar1987,Edgar1994,Edgar2003}, the constraint in Eq. (\ref{disproof2}) is sufficient  to refute the conjecture that the Riemann tensor can be derived from a tensor potential in general in four spacetime dimensions. 

\section{Concluding remarks}

A common point of view regarding the interpretation of the Lanczos' spintensor $H_{abc}$ is that it is as a local tensor potential for the Weyl tensor and discusses its properties  in parallel to the gauge theories in which the field strength is derived from a gauge potential.  
As the simplest example, the Faraday 2-form $F$ is generated from a local gauge potential $A$, by the exterior derivative $F=dA$. Such an analogy is discussed in detail by Dolan and Muratori \cite{dolan-muratori}. They also discussed the wave equation satisfied by Lanczos potential using both tensorial and spinor approaches that wwere discussed by a number of authors \cite{cl_2,odonnell,dolan-muratori,illge,dolan-kim} previously. One can also extend the discuss above to the wave equation satisfied by the Lanczos potential using a generalized Laplacian and de Rham  operators \cite{bini} acting on tensor-valued differential forms.

A discussion of Lanczos' potential in relation to a given Petrov type in a general context, in the manner discussed, for example, in Ref. \cite{castillo}, requires the formulation of the Weyl-Lanczos relation relative to a complex null coframe, thus the expression derived above (\ref{main-result}) can have further application in this regard as well. These investigations may find a considerable use in a further study on the relation between the algebraic classification of the
Weyl tensor, namely the Petrov classification based on the principle null directions, and the Lanczos potential.

A general method  for solving Weyl-Lanczos relations  by using exterior algebra was discussed by Dolan and Gerber \cite{dolan-gerber}
explicitly presenting some singular solutions for the Kasner and the G\"odel spacetimes. In connection with the work of Dolan and Gerber, 
the Weyl-Lanczos relation was already expressed in terms of differential forms which may  be helpful in a discussion for the construction of further explicit expressions.

\section*{Appendix: A derivation of the duality relations}\label{appendix}

As the following derivations explicitly indicate, the relations between left and right duals given in Eqs. (\ref{duality-rel1})-(\ref{duality-rel3}) are peculiar to four spacetime dimensions.

The inner product of two 2-forms provided by the Hodge dual is by definition symmetric in 2-forms: 
\be
\theta^{ab}\wdg *C^{cd}
=
C^{cd}\wdg *\theta^{ab}.
\ee
Using the property that $C^{ab}$ is traceless, by applying the contraction $i_di_c$ to this equation one finds
\be
\delta^{ab}_{cd}*C^{cd}
=
C^{cd}\wdg *\theta^{ab}_{\fant{ab}cd}
\ee
where $\delta^{ab}_{cd}=\delta^{a}_{c}\delta^{b}_{d}-\delta^{a}_{d}\delta^{b}_{c}$ stands for the generalized Kronecker delta symbol. 
This relation immediately yields $*C^{ab}=C^{ab}*$. 

In the same manner, by contracting the inner product
\be
\theta^{ab}\wdg *D^{cd}
=
D^{cd}\wdg *\theta^{ab}
\ee
one finds
\be
\delta^{ab}_{cd}*D^{cd}
=
S^c\wdg *\theta^{ab}_{\fant{ab}c}
+
D^{cd}\wdg *\theta^{ab}_{\fant{ab}cd}
\ee
using $i_a D^{ab}=S^b$ and $i_bi_aD^{ab}=0$. On the other hand, one can show that
$S^c\wdg *\theta^{ab}_{\fant{ab}c}=2D^{ab}$ for the first term on the right-hand side, and consequently one obtains $*D^{ab}=-D^{ab}*$.

For the coframe  basis 2-forms, one has
\be
*\theta^{ab}
=
\frac{1}{2}\epsilon^{ab}_{\fant{ab}cd}\theta^{cd}
=
\theta^{ab}*
\ee
by definition of left- and right-duals.

%
% BibTeX users please use
% \bibliographystyle{}
% \bibliography{}
%
% Non-BibTeX users please use

\end{document}